\documentclass[12pt]{article}
\usepackage{cite}
\usepackage{graphicx}
\usepackage{amssymb}
\usepackage{amsfonts}
\usepackage{amsthm}
\usepackage{amsfonts}

\bibstyle{ams}

\newcommand \no {\noindent}
\newcommand \ra {\rightarrow}

\newcommand{\ba}[1]{\begin{array}{#1}}
\newcommand{\ea}{\end{array}}

\newcommand{\be}{\begin{equation}}
\newcommand{\ee}{\end{equation}}
\newcommand{\bea}{\begin{eqnarray}}
\newcommand{\eea}{\end{eqnarray}}
\newcommand{\beann}{\begin{eqnarray*}}
\newcommand{\eeann}{\end{eqnarray*}}

\newcommand{\X}{{\cal X}}
\newcommand{\D}{{\cal D}}

\newcommand{\s}{{\sigma}}
\newcommand{\bbeta}{{\gamma}}
\newcommand{\deltah}{{h}}
\newcommand{\rfunc}{{\rho}}

\def\reff#1{(\ref{#1})}
\newtheorem{proposition}{Proposition}
\newtheorem{lemma}{Lemma}
\newtheorem{theorem}{Theorem}

\begin{document}

\title{Absence of renormalization group pathologies in some critical
  Dyson-Ising ferromagnets }

\author{Tom Kennedy
\\Department of Mathematics
\\University of Arizona
\\Tucson, AZ 85721
\\ email: tgk@math.arizona.edu
}

\maketitle 

\begin{abstract}
  The Dyson-Ising ferromagnet is a one-dimensional Ising model with a power
  law interaction. When the power is between $-1$ and $-2$, the model has a
  phase transition. Van Enter and Le Ny proved that at sufficiently
  low temperatures the decimation renormalization group transformation is not
  defined in the sense that the renormalized measure is not a Gibbs measure.
  We consider a modified model in which the nearest neighbor couplings are
  much larger than the other couplings. For a family of Hamiltonians which
  includes critical cases, we prove that the first step of the renormalization
  group transformation can be rigorously defined for majority rule and
  decimation.
\end{abstract}

\section{Introduction}

We consider a one-dimensional ferromagnetic Ising model with a power law
interaction. The standard choice for the Hamiltonian is
\bea
  H =  - \sum_{i<j} \frac{\s_i \s_j}{|i-j|^\alpha} 
\label{ham}
\eea
If the power satisfies $1 < \alpha \le 2$  the model has a phase transition.
Van Enter and Le Ny proved that when the inverse temperature is
sufficiently large, the decimation renormalization group transformation
is not defined.
In this paper we will consider a slightly modified model and prove that
the first step of the renormalization group transformation is defined in
a region of the parameters that includes critical points. Our method applies
to both decimation and the majority rule transformations.

For the Hamiltonian \reff{ham}, 
the existence of a phase transition was conjectured by 
Kac and Thompson for $1 < \alpha \le 2$ \cite{kac1969critical}.
The absence of a phase transition for $\alpha>2$ was proved by Ruelle
\cite{ruelle1968statistical}.
Dyson proved the existence of long-range order at low temperatures for 
$1< \alpha < 2$ by comparison with a hierarchical model \cite{dyson1971ising}.
Long-range order for the case of $\alpha=2$  was proved by Fr{\"o}hlich
and Spencer \cite{frohlich1982phase}.
Further properties for this case were proved in 
\cite{imbrie1982decay, imbrie1988intermediate,aizenman1988discontinuity}.
Long-range order at low temperatures has also been proved using infrared bounds
by Fr{\"o}hlich, Israel, Lieb and Simon \cite{frohlich1978phase} for 
$1 < \alpha < 2$. Non-rigorous renormalization group treatments of the
model can be found in \cite{cannas1995one,PhysRevLett.29.917,cardy1981one,
khoo2001mean}.
Bleher carried out a rigorous  renormalization group treatment of
Dyson's hierarchical model \cite{bleher1973investigation}.

Van Enter and Le Ny \cite{van2017decimation} considered the decimation
transformation with a scale factor of $2$ for this model.
(So every other spin is decimated.)
The renormalized measure is always defined; the non-trivial question is
whether it is a Gibbs measure. They prove it is not Gibbsian if the
temperature is sufficently low by exhibiting a point of essential
discontinuity for the conditional probabilities for the decimated Gibbs
measures. We refer to the spins that are fixed in the decimation
transformation as the
block spins and the spins that are not fixed as the original spins. 
The ``bad'' block spin configuration which drives their result is
the block spin configuration which alternates between $+$ and $-$.
The interactions between these fixed block spins and a single original
spin cancel exactly. What is left is a power law interaction between the
original spins. Because the lattice spacing between adjacent original spins is
$2$, this interaction is weaker than the original interaction by a factor
of $2^\alpha$. But if $\beta$ is large enough this conditioned system will
still be in the low temperature phase. 
It is important to note that this argument does not work if the original
system is near the critical temperature - the region where one is
most interested in applying the renormalization group transformation.
The method used in  \cite{van2017decimation} to prove the renormalized measure
is not Gibbsian was developed by van Enter, Fern{\'a}ndez and Sokal
\cite{van1993regularity}.

In this paper we study a slightly modified model. We will take the Gibbs
measure to be weighted by $e^{-H}$ rather than $e^{-\beta H}$ and include
parameters in $H$ that play the role of inverse temperature. 
Our Hamiltonian is 
  \bea
  H = - \bbeta \sum_i \s_i \s_{i+1} 
  - \epsilon \sum_{i<j-1} \frac{\s_i \s_j}{|i-j|^\alpha} 
  \label{myham}
  \eea
where $\bbeta,\epsilon >0$.  
The usual model at inverse temperature $\beta$ is obtained by setting
$\bbeta=\epsilon=\beta$. 
We focus our attention on the case of small $\epsilon$. Standard methods
prove there is no long-range order if both $\bbeta$ and $\epsilon$ are
sufficiently small.
For $1<\alpha<2$ existing methods can be used to prove that
given an $\epsilon>0$ there is a $\bbeta_0(\epsilon,\alpha)$ such
that there is long-range order if $\bbeta \ge \bbeta_0(\epsilon,\alpha)$.
We will sketch a proof of this using infrared bounds in section \ref{lro_sect}.
Here we show that the original non-rigorous energy-entropy argument
for long-range order applies in this case.
We impose $+$ boundary conditions on the two ends.
Then we consider a segment of $-$ spins of length $L$ which contains
the origin.
If there are no other $-$ spins in the configuration, then the energy cost
compared to the all $+$ configuation is essentially
$4 \bbeta + c \epsilon L^{2-\alpha}$
where $c$ is a constant that depends on $\alpha$. 
The number of such segments is $L$, so summing we have
\bea
\exp(-4 \bbeta) \sum_{L=1}^\infty L \, \exp(- c \epsilon L^{2-\alpha}) 
\eea
The key point is that this sum is finite for all $\epsilon>0$ if $\alpha<2$.
So given $\epsilon$ we can take $\bbeta$ to be sufficiently large to make the
above quantity as small as we like. Of course this argument is a gross
over-simplification since it
ignores the effect of other islands of $-$ spins which makes estimating
the energy cost of a contour quite difficult.
In fact, for some configurations the
energy cost of the island containing the origin need not behave as
$L^{2-\alpha}$. A much more elaborate definition of
contours would be needed to turn this into a proof.
Contour based Peierls arguments for long-range order which are based on the
ideas of \cite{frohlich1982phase} can be found in
\cite{bissacot2018contour,cassandro2005geometry,littin2017quasi}.

The main goal of this paper is to prove that there is a small $\epsilon_0$
such that for $\epsilon < \epsilon_0$ and for all $\bbeta>0$ the first step
of the RG map is defined.
Note that the parameter region in which our result holds includes
critical points. Proving that the first step of the RG map is defined
is usually easier than proving that subsequent steps are defined since the
renormalized Hamiltonian that must be considered in the subsequent steps is
considerably more complicated than the original Hamiltonian. That is the
case here as well.

Our result applies to both decimation and the majority rule transformations.
Standard expansion methods can be used to prove the result for decimation.
We provide a sketch of the idea.
We denote the original spins by $\s_i$ and denote the block spins by $s_i$.
We use blocks with two spins. So there is one block spin for every two 
original spins. The decimation transformation sets
$\s_{2i}=s_i$ for all i. In other words, the even spins in the original system
are frozen to the block spin values and the odd spins in the original system
are summed out.
The renormalized Hamiltonian $H^\prime$ is given by
\bea
\exp(-H^\prime(s)) = \sum_{\s:odd} \exp( \sum_{i:odd} \, h_i \s_i
+ \epsilon \sum_{i<j:odd} \frac{\s_i \s_j}{|i-j|^\alpha})
\eea
where each $h_i$ is a function
of the block spins. Note that $h_{2i+1}= \bbeta (s_i+s_{i+1}) + O(\epsilon)$
where the $O(\epsilon)$ term is a function of all the block spins.
In other words, the original spins which are not fixed to a block spin value
have a site-dependent magnetic field which depends on the block spins, and
there is a weak pairwise interaction between these original spins.
So we can do a high temperature expansion to obtain
a convergent expansion for $H^\prime$. 

\section{Proof of existence of first step of RG}

We study the Hamiltonian 
\bea
  H = - \bbeta \sum_i \s_i \s_{i+1}
  - \epsilon \sum_{i<j} J_{ij} \s_i \s_j
\label{hammain}
\eea
where $\epsilon$ is small and $\bbeta$ is large.
We are primarily interested in $J_{ij}=|i-j|^{-\alpha}$ for $i<j-1$,
but our method works for any translationally invariant $J_{ij}$ such that
$\sum_j |J_{0j}| <\infty$. 
We use blocks with two spins. We denote the original spins by $\s_i$
and denote the block spins by $s_i$.
Spins $\s_{2i},\s_{2i+1}$ are in the same block, and the
block spin for that block is $s_i$.
We only consider renormalization group kernels of the form
\beann
K(\s,s)= \prod_i k(\s_{2i},\s_{2i+1};s_i)
\eeann
The renormalized Hamiltonian $H^\prime(s)$ is formally defined by
\bea
\exp(-H^\prime(s)) = \sum_\s K(\s,s) \, \exp(-H(\s))
\eea
We require that the kernel satisfy
\bea
\sum_s K(\s,s) = 1, \quad \forall \s
\eea
This condition implies that the partition function is preserved by the
renormalization group transformation, i.e.,
\bea
\sum_s \exp(-H^\prime(s)) = \sum_\s \exp(-H(\s))
\eea
For the first part of this section the kernel is general. Later we will
consider specific kernels such as majority rule and decimation.

We will use a transfer matrix approach. 
Suppose we have summed out the spins $\s_i$ for $i<0$. 
The result depends on the block spins $s_i$ with $i<0$ and on the 
original spins $\s_i$ with $i \ge 0$. We can write it as
\beann
&& \sum_{\s_i:i<0} K(\s,s) \, \exp(-H(\s)) = 
\prod_{i \ge 0} k(\s_{2i},\s_{2i+1};s_i) \\
&& \exp[\sum_{X,Y: X \ge 0,Y < 0} c(X,Y) \s(X) s(Y)
 + \bbeta \sum_{i \ge 0} \s_i \s_{i+1} 
 + \epsilon \sum_{0 \le i<j} J_{ij} \s_i \s_j + E]
\eeann
The sum on $X$ is over finite subsets of the sites in the original chain and
the sum on $Y$ is over finite subsets of the sites in the renormalized chain.
The notation $X \ge 0$ means that $X$ only contains non-negative sites, and
the notation $Y < 0$ means that $Y$ only contains negative block sites.
The term $E$ does not depend on any $s_i$ or $\s_i$. We do not really
care about this term since it only contributes a constant to the renormalized
Hamiltonian $H^\prime$. 

Now consider what happens when we sum over the spins $\s_0,\s_1$. 
The resulting function will be the same as the 
previous one except that all $\s_i$ are shifted by two sites and the $s_i$
by one site. (The constant $E$ will also change.) So we have 
\beann
&&
\sum_{\s_0,\s_1} \, \prod_{i \ge 0} k(\s_{2i},\s_{2i+1};s_i) 
\, \exp[\sum_{X,Y: X \ge 0,Y<0} c(X,Y) \s(X) s(Y)
 + \bbeta \sum_{i \ge 0} \s_i \s_{i+1} 
 + \epsilon \sum_{0 \le i<j} J_{ij} \s_i \s_j +E ] \\
&&= 
\prod_{i \ge 0} k(\s_{2i+2},\s_{2i+3};s_{i+1}) 
\exp[\sum_{X,Y: X \ge 0, Y<0} c(X,Y) \s(X+2) s(Y+1)
 + \bbeta \sum_{i \ge 0} \s_{i+2} \s_{i+3} \\
&&
 + \epsilon \sum_{0 \le i<j} J_{ij} \s_{i+2} \s_{j+2} +E^\prime]
\eeann
We use $X+k$ to denote $\{i+k:i \in X\}$, i.e.,  
the set of sites in $X$ with each site shifted by $k$. 
After some obvious cancellations and re-indexing of the terms in the right side,
this simplifies to 
\beann
&&
\sum_{\s_0,\s_1} \, k(\s_0,\s_1;s_0) \, \exp[\sum_{X,Y: X \ge 0, Y<0} c(X,Y) \s(X) s(Y)
  + \bbeta \sum_{i=0,1} \s_i \s_{i+1} 
  + \epsilon \sum_{i=0,1} \sum_{j>0} J_{ij} \s_i \s_j
] \\
= 
&& \exp[\sum_{X,Y: X \ge 2, Y \le 0} c(X-2,Y-1) \s(X) s(Y) ]
\eeann
Let $\X$ be the collection of $X$ such that $X \ge 0$ and $X$ contains at least
one of $0,1$. Note that the $X$ not in $\X$ are precisely the $X$ with
$X \ge 2$. Pulling such terms outside the sum on $\s_0,\s_1$ we have
\bea
&&
\exp[\sum_{X,Y: X \ge 2, Y<0} c(X,Y) \s(X) s(Y)]
\sum_{\s_0,\s_1} \, k(\s_0,\s_1;s_0) \, \nonumber \\
&& \exp[\sum_{X,Y: X \in \X, Y<0} c(X,Y) \s(X) s(Y)
 + \bbeta \sum_{i=0,1} \s_i \s_{i+1}
  + \epsilon \sum_{i=0,1} \sum_{j>0} J_{ij} \s_i \s_j] \nonumber \\
= 
&& \exp[\sum_{X,Y: X \ge 2, Y \le 0} c(X-2,Y-1) \s(X) s(Y) ]
\label{prefpe}
\eea

For a general Hamiltonian of the form
\bea
H = \sum_X h(X) \s(X)
\eea
we define
\bea
\hat{H} = \sum_{X \in \X} h(X) \s(X)
\eea
So for our Hamiltonian
\bea
\hat{H} = \bbeta \sum_{i=0,1} \s_i \s_{i+1}
  + \epsilon \sum_{i=0,1} \sum_{j>0} J_{ij} \s_i \s_j 
\eea
This is the part of the Hamiltonian that appears in \reff{prefpe}.
We let $c$ denote the collection $\{c(X,Y): X \in \X, Y<0 \}$. 
Note that $\hat{H}$ appears in \reff{prefpe} in exactly the same way that $c$
does. Now we define $f(c,X,Y)$ for $X \ge 2, Y \le 0$ by
\bea
&& \sum_{\s_0,\s_1} \, k(\s_0,\s_1;s_0) \, 
 \exp[\sum_{X,Y: X \in \X,Y < 0} c(X,Y) \s(X) s(Y)] \nonumber \\
&& =
\exp[\sum_{X,Y: X \ge 2, Y \le 0} f(c,X,Y) \s(X) s(Y)]  
\label{fdef}
\eea
Note that the Hamiltonian does not appear in the above. 
We will include the Hamitonian by replacing the $c$ in $f(c,X,Y)$ by
$c+\hat{H}$. In particular equation \reff{prefpe} becomes
\bea
&& \sum_{X,Y: X \ge 2, Y<0} c(X,Y) \s(X) s(Y) +
\sum_{X,Y: X \ge 2, Y \le 0} f(c+\hat{H},X,Y) \s(X) s(Y)   \nonumber \\
&=& 
\sum_{X,Y: X \ge 2, Y \le 0} c(X-2,Y-1) \s(X) s(Y)
\label{eqg}
\eea

Equation \reff{eqg} implies that 
\beann
c(X-2,Y-1) &=& c(X,Y) + f(c+\hat{H},X,Y), \quad if \quad X \ge 2, \, Y < 0 \\
c(X-2,Y-1) &=& f(c+\hat{H},X,Y), \quad if \quad X \ge 2, \, Y \le 0, \, 0 \in Y
\eeann
This is equivalent to 
\beann
c(X,Y) &=& c(X+2,Y+1) + f(c+\hat{H},X+2,Y+1), \quad if \quad X \ge 0, \,
Y < -1 \\
c(X,Y) &=& f(c+\hat{H},X+2,Y+1), \quad if \quad X \ge 0, \, Y \le -1, \,
-1 \in Y
\eeann
If $Y$ is non-empty, then by iterating the above we find 
\beann
c(X,Y) &=& \sum_{k=1}^n f(c+\hat{H},X+2k,Y+k), \quad if \quad X \ge 0, \,
Y < 0
\eeann
where $n$ is the largest (least negative) site in $Y$. 
If $Y=\emptyset$ then we have 
\beann
c(X,\emptyset) &=& \sum_{k=1}^n f(c+\hat{H},X+2k,\emptyset)
+ c(X+2n,\emptyset), \quad if \quad X \ge 0
\eeann
Assuming that $c(X+2n,\emptyset)$ converges to $0$ as $n \rightarrow \infty$,
we have 
\beann
c(X,\emptyset) &=& \sum_{k=1}^\infty f(c+\hat{H},X+2k,\emptyset) 
\quad if \quad X \ge 0
\eeann
If we define $f(c+\hat{H},X,Y)$ to be zero when $Y$ does not satisfy $Y \le 0$,
then we have 
\bea
c(X,Y) &=& \sum_{k=1}^\infty f(c+\hat{H},X+2k,Y+k) 
\quad if \quad X \ge 0, \, Y<0
\label{fpe}
\eea
for both the case of non-empty $Y$ and the case of empty $Y$. 

We define for $X \in \X$ and $Y<0$ 
\bea
F(c,X,Y) = \sum_{k=1}^\infty f(c,X+2k,Y+k)
\label{fsum}
\eea
Then \reff{fpe} is a fixed point equation for $c$, 
\bea
F(c+\hat{H},X,Y)=c(X,Y), \quad X \in \X, Y < 0
\label{fixpt}
\eea
The function $F(c,X,Y)$ is defined for all $X$ with $X \ge 0$. But in the
fixed point equation we only use it for $X$ with $X \in \X$, i.e., $X$ contains
at least one of $0,1$. 

Note that $f(c+\hat{H},X,Y)$ is defined when $X=\emptyset$ and $Y \le 0$,
but these functions do not appear in the fixed point equation above. 
They are terms in the renormalized Hamiltonian $H^\prime$. 
The contribution to $H^\prime$ from summing out $\s_0,\s_1$ is
\beann
\sum_{Y: Y \le 0} f(c+\hat{H},\emptyset,Y) \, s(Y)
\eeann
To obtain $H^\prime$ we must sum this over translations with respect to
the block lattice. Recall that we define $f(c+\hat{H},\emptyset,Y)$ to
be zero when $Y$ does not satisfy $Y \le 0$. So we have
\beann
H^\prime = \sum_{k= -\infty}^\infty \sum_Y
f(c+\hat{H},\emptyset,Y) \, s(Y+k)
= \sum_Y h^\prime(c+\hat{H},Y) s(Y) 
\eeann
where the coefficient of $s(Y)$ is given by 
\bea
h^\prime(c+\hat{H},Y) = \sum_{k= -\infty}^\infty f(c+\hat{H},\emptyset,Y-k)
= \sum_{k= -\infty}^\infty f(c+\hat{H},\emptyset,Y+k)
\label{hdef}
\eea
At the moment this is a purely formal expression. (We do not know that this
infinite series converges.) We will make the definition of the renormalized
Hamiltonian rigorous later. 

We will prove that Eq. \reff{fixpt} has a fixed point by constructing an 
approximate fixed point and showing the map is a contraction in a sufficiently
large neighborhood of it.
Let $g(X,Y)$ be a function where $X$ ranges over finite subsets of the original
lattice and $Y$ ranges over finite subsets of the block lattice.
We define the norm of $g$ to be 
\bea
||g|| = \sum_{X,Y} |g(X,Y)| \, \exp(\mu |X|)
\label{gnorm}
\eea 
where $|X|$ is the cardinality of $X$, and $\mu \ge 0$.  
We will only allow functions $g$ with finite $||g||$. 
The functions $c$ that occur in the fixed point equation have an
additional property: 
$c(X,Y) \ne 0$ only for $X \in \X$ and $Y<0$.  
The Banach space in which we look for a solution to the fixed-point 
equation is the set of $c$'s with this property and finite $||c||$. 
We will show that $F(c)$ and its Jacobian $DF(c)$ are defined and 
continuous on an open subset of this Banach space. 
We defined the norm \reff{gnorm} for functions on all $X,Y$ for later use. 

Given such a function $g(X,Y)$ with $||g||<\infty$, we
can define a function of $\s,s$ by
\bea
g(\s,s)= \sum_{X,Y} g(X,Y) \s(X) s(Y)
\label{gexp}
\eea
So we can think of the norm as a norm on functions of $\s,s$ of the form
\reff{gexp}.

Given two functions $g_1(X,Y),g_2(Y,Y)$ we can multiply the functions
$g_1(\s,s)$ and $g_2(\s,s)$. 
Since $\s_i^2=1$, we have $\s(A) \s(B) = \s(A \Delta B)$ where
the symmetric difference is defined by 
$A \Delta B = A \cup B \setminus (A \cap B)$.
Similarly, $s(A) s(B) = s(A \Delta B)$.
Thus
\beann
g_1(\s,s) \, g_2(\s,s)
&=& \sum_{X_1,X_2,Y_1,Y_2} g_1(X_1,Y_1) g_2(X_2,Y_2)
\s(X_1 \Delta X_2) s(Y_1 \Delta Y_2)\\
&=& \sum_{X,Y} \s(X) s(Y) \sum_{X_1,X_2,Y_1,Y_2:X_1 \Delta X_2=X,Y_1 \Delta Y_2=Y}
g_1(X_1,Y_1) g_2(X_2,Y_2)
\eeann 
so
\beann
||g_1 g_2|| &\le& \sum_{X_1,X_2,Y_1,Y_2} |g_1(X_1,Y_1) g_2(X_2,Y_2)|
\exp(\mu |Y_1 \Delta Y_2|) \\
&\le& \sum_{X_1,X_2,Y_1,Y_2} |g_1(X_1,Y_1) g_2(X_2,Y_2)|
\exp(\mu (|Y_1|+|Y_2|) = ||g_1|| \, ||g_2||
\eeann 
where we have used $|Y_1 \Delta Y_2| \le |Y_1| + |Y_2|$.
We will use this Banach algebra property extensively.

Suppose we have a function $g(\s,s)$ which has finite support in the sense
that it only depends on finitely many $\s_i$ and $s_i$. Then it can be
written in the form \reff{gexp}. To compute the coefficients $g(X,Y)$,
first consider
\beann
\sum_{\s,s} \s(X) s(Y) \s(V) s(W) = 
\sum_{\s,s} \s(X \Delta V) s(Y \Delta W)
\eeann
where the sum is only over the $\s_i$ and $s_i$ in the support of $g$. 
This sum is $0$ unless $X=V$ and $Y=W$, in which case it equals the number 
of terms in the sum. Letting $N$ denote the number of terms in the sum, 
\bea
g(X,Y) = {1 \over N} \sum_{\s,s} \s(X) s(Y) \, g(\s,s)
\label{inversion}
\eea

If $c$ has finite support, then eqs.\reff{fdef} and \reff{inversion} imply 
\bea
f(c,V,W) = {1 \over N} \sum_{\s,s} \s(V) s(W) 
 \ln [
 \sum_{\s_0,\s_1} \, k(\s_0,\s_1;s_0) \, 
\exp(\sum_{X,Y: X \in \X, Y<0} c(X,Y) \s(X) s(Y))] 
\label{fdef_inv}
\eea
If we consider $f(c+\hat{H},V,W)$ then finite support means that $\hat{H}$
must have finite support as well. 
We will initially work in the case of 
finite support and then extend our definitions and bounds to an open 
subset of $c$ and the full Hamiltonian. 

We will extend the definition of $F$ by obtaining bounds on its derivative. 
Let $DF(c)$ denote the Jacobian of $F$ at $c$. Its matrix elements are 
$\partial F(c,V,W) / \partial c(X,Y)$ 
The norm we are using is a weighted $l^1$ 
norm, so the operator norm of $DF(c)$ is bounded by 
\beann
||DF(c)|| \le \sup_{X \in \X,Y<0} \sum_{V \in \X, W<0} 
\left| {\partial F(c,V,W) \over \partial c(X,Y)} \right| \exp(\mu |V| - \mu|X|)
\eeann
As before $\X$ denotes the set of finite subsets $X$ with
$X \ge 0$ and at least one of $0,1$ is in $X$. 
From \reff{fsum} we have 
\beann
{\partial F(c,V,W) \over \partial c(X,Y)} = \sum_{k=1}^\infty 
{\partial f(c,V+2k,W+k) \over \partial c(X,Y)} 
\eeann

Let $g(\s,s)$ be a function of the spins $\s_i$ with $i \ge 0$ and the block
spins $s_i$ with $i<0$. We define 
\bea
<g>_c = {1 \over Z}  \sum_{\s_0, \s_1} \, g(\s,s) \, k(\s_0,\s_1;s_0) \, 
 \exp(\sum_{X,Y: X \in \X,Y<0} c(X,Y) \s(X) s(Y) )
\label{expdef}
\eea
where $Z$ is defined by $<1>_c=1$.
Note that the Hamiltonian does not appear in the definition of $<g>_c$.
As before, we include the Hamiltonian by considering $<g>_{c+\hat{H}}$.
Note that $<g>_c$ is a function of $\s_i$ with $i \ge 2$ and $s_i$ with
$i \le 0$. 
At this point $<g>_c$ is only defined if $c$ and $g$ have finite support, and so
$<g>_{c+\hat{H}}$ is only defined if $c$, $\hat{H}$ and $g$ have finite support.

If $c, \hat{H}$ and $g$ have finite support, then
eq. \reff{fdef} implies 
\beann
{\partial f(c,V+2k,W+k) \over \partial c(X,Y)} = {1 \over N} \sum_{\s,s} 
\s(V+2k) \, s(W+k) \, <\s(X) s(Y)>_c
\eeann
Since $X$ contains at least one of $0,1$, the term $\s(X)$ contains a factor
of either $\s_0$, $\s_1$ or $\s_0 \s_1$. The rest of $\s(X)$ can be factored
out of the expectation. 
So we need to consider $<\s_0>_c$, $<\s_1>_c$ and $<\s_0 \s_1>_c$. 
They have expansions of the form 
\beann
<\s(A)>_c = \sum_{U,T: U \ge 2, T \le 0} d(c,A,U,T) \s(U) s(T)
\eeann
where $A$ is $\{0\}$, $\{1\}$ or $\{0,1\}$.
Note that this sum includes the term with $U=\emptyset$ and $T=\emptyset$. 
The coefficients $d(c,A,U,T)$ are given by 
\beann
d(c,A,U,T) = {1 \over N} \sum_{\s,s} \s(U) s(T) <\s(A)>_c
\eeann
For $c$ with finite support
$d(c,A,U,T)$ is nonzero for only finitely many $U,T$. 

Let $A=X \cap \{0,1\}$. So $\s(X)= \s(A) \s(X \setminus A)$. 
Using $\s(P) \s(Q) = \s(P \Delta Q)$ we have
\beann
    {\partial f(c,V+2k,W+k) \over \partial c(X,Y)} &=&
    {1 \over N} \sum_{\s,s} \s((V+2k)\Delta(X\setminus A)) \,
    s((W+k) \Delta Y)) \, <\s(A)>_c \\
    &=&
    d(c,A,(V+2k) \Delta (X \setminus A),(W+k) \Delta Y)
\eeann
Thus 
\beann
{\partial F(c,V,W) \over \partial c(X,Y)} = \sum_{k=1}^\infty
    d(c,A,(V+2k) \Delta (X \setminus A),(W+k) \Delta Y)
\eeann

So for any $X,Y$ with $X \in \X$, $Y <0$, 
\bea
&& 
\sum_{V \in \X, W<0} 
\left| {\partial F(c,V,W) \over \partial c(X,Y)} \right| \,
\exp(\mu |V| - \mu|X|) \\
&\le&
\sum_{V \in \X,W<0} \sum_{k=1}^\infty
|d(c,A,(V+2k) \Delta (X \setminus A),(W+k) \Delta Y)|
\, \exp(\mu |V| - \mu|X|)
\label{dbound}
\eea
It is trivial to check that for any sets $P,Q$ we have
$|P| \le |P \Delta Q| +|Q|$. So we have
\beann
|V|=|V+2k| \le |(V+2k) \Delta (X \setminus A)| + |X \setminus A|
\eeann
We have $|X \setminus A| = |X|-|A|$.
Thus \reff{dbound} is bounded by 
\beann 
\le  
e^{-\mu |A|} \sum_{V \in \X,W<0} \sum_{k=1}^\infty
|d(c,A,(V+2k) \Delta (X \setminus A),(W+k) \Delta Y)|
\, \exp(\mu |(V+2k) \Delta (X \setminus A)|)
\eeann
We order this sum as 
\beann
e^{-\mu |A|} \sum_{V \in \X} \sum_{k=1}^\infty \sum_{W<0} 
|d(c,A,(V+2k) \Delta (X \setminus A),(W+k) \Delta Y)|
\, \exp(\mu |(V+2k) \Delta (X \setminus A)|)
\eeann
Recall that $Y$ is fixed here. Now consider a fixed $k$. 
The map $W \rightarrow (W+k) \Delta Y$ is one-to-one. 
As $W$ ranges over all finite subsets with $W<0$, $(W+k) \Delta Y$
will include some subsets which have a site $\ge 0$.
But the coefficient $d()$ vanishes for
these cases. So we get an upper bound on the above by replacing 
$(W+k) \Delta Y$ by just $W$ and summing over $W$ subject to $W \le 0$:
\beann
\le && e^{-\mu |A|} \sum_{V \in \X} \sum_{k=1}^\infty \sum_{W \le 0} 
|d(c,A,(V+2k) \Delta (X \setminus A),W)| 
 \exp(\mu |(V+2k) \Delta (X \setminus A)|)
\eeann
Now $V \in \X$ implies at least one of $0,1$ is in $V$.
So as we sum over $V \in \X$ and $k$,
$V+2k$ will range over all subsets $ \ge 2$. Since $X$ is fixed,
$(V+2k) \Delta (X \setminus A)$ will also range over all subsets $\ge 2$. 
So the above equals
\beann
e^{-\mu|A|} \sum_{U \ge 2} \sum_{W\le 0} 
|d(c,A,U,W)| \, \exp(\mu |U|)
\eeann
Note that this equals $e^{-\mu|A|} ||<\s(A)>_c||$. Define
\bea
\D(c) = \max_A e^{-\mu |A|} \, ||<\s(A)>_c||
=\max_A e^{-\mu |A|} \, 
\sum_{U \ge 2,T \le 0} |d(c,A,U,T)| \, e^{\mu |U|} 
\label{ddef}
\eea
where the max is over non-empty subsets $A$ of $\{0,1\}$.
Thus we have proved
\begin{lemma}
\bea
||DF(c)|| \le \D(c)
\eea
\label{lema}
\end{lemma}

We work in the following open subset of the Banach space:
\bea
O = \{ c: c=c_0 + \delta, c_0 \, has \, finite \, support,
||\delta||<\ln 2, \D(c_0) + \rfunc(||\delta||)<1\}
\label{openset}
\eea
where $\rfunc(r)=2(e^r-1)/(2-e^r)$. 
Note that $\rfunc(r) \ra 0$ as $r \ra 0$. 

\begin{lemma} 
Let $g$ be a function of $\s,s$ with finite support. Let $c$ have finite
support with $\D(c)<1$. Define $<g>_c$ by \reff{expdef}.
Then 
\bea
||<g>_c|| \le ||g||
\label{gbound}
\eea
If $c_1,c_2$ have finite support with $\D(c_i)<1$, $i=1,2$ and 
$||c_1-c_2||<\ln 2$, then 
\bea
||<g>_{c_1} - <g>_{c_2}|| \le 
\rfunc(||c_1-c_2||) \, ||g||
\label{gdifbound}
\eea
and 
\bea 
| \D(c_1) -  \D(c_2) | 
\le \rfunc(||c_1-c_2||)
\label{dfdifbound}
\eea
The set of $c$ with finite support and $\D(c)<1$ is dense in $O$. 
Thus there is a unique continuous extension of the definition of $<g>_c$ 
to all $c \in O$ and all $g$ with $||g|| < \infty$. Furthermore
\reff{gbound} holds for all $c \in O$ and 
\reff{gdifbound},\reff{dfdifbound} hold for all $c_1,c_2 \in O$ with 
$||c_1-c_2|| < \ln 2$.
\label{lemb}
\end{lemma}

\no {\bf Proof:}
The bound \reff{gbound} will follow immediately from this bound for the 
case of $g(\s)=\s(V) s(W)$ where $V,W$ are finite subsets. If $V$ does not
contain either of $0,1$, then $<\s(V) s(W)>_c$ is just $\s(V) s(W)$
and the bound is immediate. 
Now suppose $A=V \cap \{0,1\}$ is non-empty. Then
\beann
<\s(V) s(W)>_c &=& \s(V \setminus A) \, s(W) \, <\s(A)>_c \\
&=& \s(V \setminus A) \, s(W) \, \sum_{U \ge 2,T \le 0} d(c,A,U,T) \, \s(U)
\, s(T) \\
&=& \sum_{U \ge 2,T \le 0} d(c,A,U,T) \, \s(U \Delta (V \setminus A))
\, s(T \Delta W) 
\eeann
So
\beann
||<\s(V) s(W)>_c||
&\le& \sum_{U \ge 2,T \le 0} |d(c,A,U,T)| \, \exp(\mu |U \Delta (V \setminus A)|)
\eeann
We use
\beann
|U \Delta (V \setminus A)| \le |U| + |V \setminus A|
= |U| + |V| - |A|
\eeann
So the above is 
\beann
&\le& \sum_{U \ge 2,T} |d(c,A,U,T)| \, \exp(\mu (|U| + |V| - |A|)) \\
&\le& \D(c) \, \exp(\mu |V|) 
\eeann
Since $||\s(V)||=\exp(\mu |V|)$ and $\D(c) < 1$, the bound follows.

For the bound \reff{gdifbound}, we
denote $c_1$ by $c$ and let $\delta=c_2-c_1$ 
so $c_2=c+\delta$. Then we can express the quantity we need to bound as
\beann
<g>_{c+\delta} - <g>_c 
&=& {<g \exp(\delta)>_c \over <\exp(\delta)>_c} - <g>_c\\
&=& {<g \exp(\delta)>_c - <g>_c <\exp(\delta)>_c \over <\exp(\delta)>_c} 
\eeann
Let $k=\exp(\delta)-1$. Since we are in a Banach algebra we have
\beann
||k|| = ||\sum_{n=1}^\infty {\delta^n \over n!} || \le   
\sum_{n=1}^\infty {||\delta||^n \over n!} = \exp(||\delta||)-1
\eeann
Note that since $||\delta||<\ln 2$, $||k||<1$. 
The quantity we need to bound is 
\beann
 {<g (1 + k)>_c - <g>_c <1 + k>_c \over <1 + k >_c} 
= {<g k>_c - <g>_c <k>_c \over 1 + <k>_c} 
\eeann
Now we use \reff{gbound} and a power series expansion and the 
fact that we are in a Banach algebra to see
\beann
||(1 + <k_c>)^{-1}|| \le (1 - ||k||)^{-1}
\eeann
We use
\beann
||<gk>_c>|| &\le& ||gk|| \le ||g|| \, ||k|| \\
||<g>_c \, <k>_c>|| &\le& ||<g>_c|| \, ||<k>_c>|| \le ||g|| \, ||k|| 
\eeann
Thus
\beann
||<g>_{c+\delta} - <g>_c || 
&\le& 2 ||g|| \, ||k|| (1 - ||k||)^{-1}
\nonumber \\
&\le& 2 ||g|| \, (\exp(||\delta||) -1) (2 - \exp(||\delta||))^{-1} 
= \rfunc(||\delta||) \, ||g||
\eeann

The final bound \reff{dfdifbound} follows from \reff{gbound} 
and the definition of $\D(c)$.
\qed

\begin{lemma} 
Let $c_1,c_2$ have finite support with $||c_1-c_2||< \ln 2$.
Suppose that
\bea
\D(c_i)+\rfunc(||c_1-c_2||) \le 1
\eea
for $i=1,2$. 
Then 
\beann
||DF(c_1)-DF(c_2)|| &\le& \rfunc(||c_1 - c_2||) \nonumber \\
||F(c_1)-F(c_2)|| &\le& ||c_1 - c_2|| 
\nonumber 
\eeann
\label{lemc}
\end{lemma}

\no {\bf Proof:}
The first assertion follows immediately from lemmas \ref{lema} and \ref{lemb}.
The first assertion and the hypotheses imply
\bea
||DF(tc_1 + (1-t)c_2)|| \le 1 
\eea
for $0 \le t \le 1$. The second assertion then follows from
the first by writing $F(c_1)-F(c_2)$ as the integral of the derivative
of $F(tc_1 + (1-t)c_2)$ with respect to $t$. 
\qed

Recall that we want to solve the fixed point equation $F(c+\hat{H})=c$, but
the Hamiltonian $H$ does not appear in the lemmas above. 
Let $H_0$ be a truncation of the Hamiltonian $H$ such that $\hat{H_0}$ has
finite support. Then the above lemmas show $F(c+\hat{H_0})$ is defined
if $c+\hat{H_0}$ is in $O$, and so $F(c+\hat{H})$ is defined if
$||\hat{H}-\hat{H_0}||$ is sufficiently small. A sufficient criteria for
the existence of a solution $c$ of $F(c+\hat{H})=c$ is given in the next
theorem.

\begin{theorem}
  Let $H_0$ be a Hamiltonian such that $\hat{H_0}$ has
  finite support. 
  Let $\deltah = ||\hat{H}-\hat{H_0}||$. 
  Define $\D(c)$ by \reff{ddef}. 
  Suppose there is an approximate fixed point $c_0$
  such that $\D(c_0+\hat{H_0}) < 1$ and 
  \bea
  \inf_{r>0} \quad \frac{||F(c_0+\hat{H_0})-c_0|| + \deltah}
      {r [1-\D(c_0+\hat{H_0}) - \rfunc(r+\deltah)]} < 1
  \label{thmhyp}
  \eea
  where $\rfunc(r)=2(e^r-1)/(2-e^r)$. 
  Then there a solution to the fixed point equation $F(c+\hat{H})=c$.
\label{fpethm}
\end{theorem}

\no {\bf Proof:}
A standard argument shows there is a fixed point if there is an $r>0$
and $C < 1$ such that $||DF(c+\hat{H})|| \le C$ for $||c-c_0||\le r$ 
and $||F(c_0+\hat{H})-c_0|| \le r (1-C)$.
By the hypothesis there is an $r>0$ such that
\bea
||F(c_0+\hat{H_0})-c_0|| + \deltah < [1-\D(c_0+\hat{H_0}) -
  \rfunc(r+\deltah)] r
\eea
By lemma \ref{lemc}
\beann
||F(c_0+\hat{H})-c_0|| \le ||F(c_0+\hat{H_0})-c_0|| + \deltah
\eeann
For $c$ such that $||c-c_0|| \le r$, lemma \ref{lemc} also implies 
\bea
||DF(c+\hat{H})|| \le ||DF(c_0+\hat{H_0})|| + \rfunc(||c-c_0|| + \deltah)
\le \D(c_0+\hat{H_0}) + \rfunc(r + \deltah)
\eea
So we can take $C=\D(c_0+\hat{H_0}) + \rfunc(r + \deltah)$ and the
theorem follows.
\qed

\bigskip

The final step is to show that the renormalized Hamiltonian is defined.
The renormalized Hamiltonian is formally given by 
\beann
H^\prime= \sum_W h(c+\hat{H},W) s(W)
\eeann
where 
\bea
h(c+\hat{H},W) = \sum_{k= -\infty}^\infty f(c+\hat{H},\emptyset,W+k)
\eea
If $c+\hat{H}$ has finite support, then the series defining $h(W)$
is finite, so $h(c+\hat{H},W)$ is defined. We need to extend it to
$c+\hat{H}$ that do not have finite support. 

\begin{theorem}
  Define $O$ as in  \reff{openset}. Then $h(c+\hat{H},W)$ has a unique
  continuous extension to $c+\hat{H} \in O$. Moreover, for such $c+\hat{H}$ 
  \bea
  \sum_{W: 0 \in W} |h(c+\hat{H},W)| < \infty
  \label{flip}
  \eea
  So the change in the renormalized Hamiltonian from flipping a
  single spin is finite. 
\end{theorem}

\no {\bf Proof:}
It suffices to prove the statements in the theorem with $c+\hat{H}$ replaced
by $c$. We first show that $h(c,W)$ has a unique continuous extension from the
$c$ with finite support in $O$ to all of $O$.
For $c_0$ with finite support and $\D(c_0)<1$, let $r(c_0)$ be the
solution to $\D(c_0)+\rfunc(r(c_0))=1$.
Let $B_{r(c_0)}(c_0)$ denote the ball centered at $c_0$ with radius $r(c_0)$.
Then $O$ is the union over
$c_0$ with finite support and $\D(c_0)<1$ of $B_{r(c_0)}(c_0)$.
So it suffices to prove there is a unique
continuous extension on each of these balls. 
Suppose $c_0+\delta \in B_{r(c_0)}(c_0)$ and $\delta$ has finite support. 
Then we can write the difference $h(c_0+\delta,W)-h(c_0,W)$
as the integral of the derivative with respect to $t$ of
$h(c_0 + t\delta,W)$. The integrand can be bounded by
$||\delta||$ times the following bound on the gradient of $h$:
\bea
\sup_{0 \le t \le 1} \sup_{X \in \X, Y<0} e^{-\mu|X|}
|\frac{\partial h(c_0+t \delta,W)}{\partial c(X,Y)}|
\label{gradbound}
\eea
Let $A= X \cap \{0,1\}$. With $d()$ defined as before we have
\bea
\frac{\partial h(c+t\delta,W)}{\partial c(X,Y)}
= \sum_{k= -\infty}^\infty \, d(c_0 + t \delta,A,X \setminus A,(W+k) \Delta Y)
\label{hpartial}
\eea
For fixed $W$ and $Y$, the map $k \rightarrow (W+k) \Delta Y$ is one to one.
So \reff{gradbound} is bounded by
\bea
\max_A e^{-\mu |A|} \, \sum_{T \le 0} |d(c_0 + t \delta,A,X \setminus A,T)| 
\le \D(c_0 + t\delta) \le 1
\eea
were the last inequality follows from \reff{dfdifbound}.
Thus $|h(c_0+\delta,W)-h(c_0,W)| \le ||\delta||$. Thus 
$h(c,W)$ has a unique continuous extension to all of $B_{r(c_0)}(c_0)$. 

For the last assertion of the theorem we need to sum over $W$ which
contain $0$. However, the map from $W,k$ to $W+k$ is not one to one if
we allow all $W$ containing $0$. (Given $V$, the number of $W,k$ such
that $W+k=V$ is $|V|$.). If we restrict the sum to $W$ such that $W \ge 0$
and $0 \in W$ then the map is one to one. So the argument above shows that
\beann
  \sum_{W: 0 \in W, W \ge 0} |h(c,W)| < \infty
\eeann
This is weaker than what we want since
\beann
\sum_{W: 0 \in W} |h(c,W)| =
\sum_{W: 0 \in W, W \ge 0} |h(c,W)| |W|
\eeann
We need to modify the norm we use to obtain the stronger result. In place
of the norm \reff{gnorm} we define
\bea
||g|| = \sum_{X,Y} |g(X,Y)| \, \exp(\mu |X| + \nu |Y|)
\label{newnorm}
\eea 
where $\mu,\nu \ge 0$. Everything we have done before goes through with
this norm. If the hypotheses of theorem \reff{fpethm} hold for
some $\mu>0$ and $\nu=0$, then they hold for that $\mu$ and sufficiently
small $\nu$. This implies
\beann
\sum_{W: 0 \in W, W \ge 0} |h(c,W)| e^{\nu |W|} < \infty
\eeann
for some positive $\nu$. This implies \reff{flip}.
\qed

\subsection{Decimation}

We now consider specific RG kernels. For decimation we take
\bea
k(\s_0,\s_1;s_0) = \frac{1}{2} + \frac{1}{2} s_0 \s_1 = \delta_{s_0,\s_1}
\eea
This fixes the original spin $\s_1$ to be equal to the block spin $s_0$.
We take $H_0$ to just be the nearest neighbor part of $H$:
\bea
  H_0 = - \bbeta \sum_i \s_i \s_{i+1} 
\eea
So $h=||\hat{H}-\hat{H_0}|| = c(\alpha) \epsilon$ where
\bea
c(\alpha) =  2 \sum_{j=2}^\infty j^{-\alpha}
\eea
We take the approximate fixed point $c_0$ to just have one term:
\bea
c_0 = \bbeta s_{-1} \s_0
\eea
Then \reff{fdef} becomes
\beann
\exp[\sum_{X,Y: X \ge 2} f(c_0+\hat{H_0},X,Y) \s(X) s(Y)]
&=& \sum_{\s_0} \, \exp[\bbeta s_{-1} \s_0  + \bbeta \s_0 s_0 + \bbeta s_0 \s_2]
\\
&=& \exp(\bbeta s_0 \s_2)
\sum_{\s_0} \, \exp[\bbeta s_{-1} \s_0  + \bbeta \s_0 s_0] \\
&=& \exp(\bbeta s_0 \s_2 + c(\bbeta) s_{-1} s_0 +E)
\eeann
where $c(\bbeta)=\frac{1}{2} \ln \cosh(2 \bbeta)$ and $E$ is a constant
we do not care about.
The $\bbeta s_0 \s_2$  term contributes $\bbeta s_{-1} \s_0$ to
$F(c_0+\hat{H_0})$. The $c(\bbeta) s_{-1} s_0$ term does not contribute to
$F(c_0+\hat{H_0})$; it contributes to the renormalized Hamiltonian. 
So we have $F(c_0+\hat{H_0})=c_0$. This is reflection of the fact that
decimation is trivial for the Hamiltonian with $\epsilon=0$. 

With this approximate fixed point
\beann
<\s_0>_{c_0+\hat{H_0}} &=& \frac{d}{2} (s_0+s_{-1}) \\
<\s_1>_{c_0+\hat{H_0}} &=& s_0 \\
<\s_0 \s_1>_{c_0+\hat{H_0}} &=& \frac{d}{2} (1+ s_0 s_{-1}) \\
\eeann
where $d=\tanh(2 \bbeta)$.  So
\beann
||<\s_0>_{c_0+\hat{H_0}}|| &=& d \\
||<\s_1>_{c_0+\hat{H_0}}|| &=& 1 \\
||<\s_0 \s_1>_{c_0+\hat{H_0}}|| &=& d \\
\eeann
Noting that $d \le 1$ for all $\bbeta$, this yields
\bea
\D(c) \le e^{-\mu}
\eea
Thus if $\epsilon$ is sufficiently small we can choose $\mu>0$ so that 
\reff{thmhyp} is satisfied.

\subsection{Majority rule}

The RG kernel is 
\bea
k(\s_0,\s_1;s_0) = \frac{1}{2} + \frac{1}{4} s_0 (\s_0+\s_1)
\eea
Unlike decimation, when $\epsilon=0$ the fixed point equation is not satisfied
by a $c$ with finite support. Nonetheless, the fixed point equation is very
well-behaved in this case.
For the approximate fixed point it will suffice to only
include two terms: $\s_0 s_{-1}$ and $\s_0 s_{-2}$.
As we will see, for large $\bbeta$ the coefficient of the first term
will be of the form $\bbeta+a$ where $a$ is essentially constant,
and the coefficient of the second term will be essentially constant.
So we take the approximate fixed point to be
\beann
c_0= (\bbeta + a) \s_0 s_{-1} + b \s_0 s_{-2}
\eeann
So we consider
\beann
\sum_{\s_0,\s_1} \, k(\s_0,\s_1;s_0) \exp[\bbeta \s_0 \s_1 + \bbeta \s_1 \s_2
  + (\bbeta + a) \s_0 s_{-1} + b \s_0 s_{-2}]
\eeann
This is an even function of $\s_2, s_0, s_{-1}, s_{-2}$,
so it must be of the form 
\bea
\exp[c \s_2 s_0 + d \s_2 s_{-1} + e \s_2 s_{-2} + f \s_2 s_0 s_{-1} s_{-2} 
  + g s_0 s_{-1} + h s_0 s_{-2} + k s_{-1} s_{-2} + E]
\eea
Note that terms which only contain block spins do not contribute to
$F(c_0+\hat{H_0})$. They only contribute to the renormalized Hamiltonian. 
We have 
\beann
F(c_0+\hat{H_0}) = c \s_0 s_{-1} + d \s_0 s_{-2} + e \s_0 s_{-3}
+ f \s_0 s_{-1} s_{-2} s_{-3} 
\eeann
So we have
\bea
F(c_0+\hat{H_0}) - c_0 =
(c - \bbeta -a ) \s_0 s_{-1}
+ (d-b) \s_0 s_{-2} + e \s_0 s_{-3}
+ f \s_0 s_{-1} s_{-2} s_{-3}
\label{fpeerr}
\eea
We find after some algebra that
\beann
c &=& \bbeta + (-2*\ln(1.5)+x-y+z-w)/8  + O(e^{-\bbeta}) \\
d &=& (-2*\ln(1.5)-x+y-z+w)/8 + O(e^{-\bbeta}) \\
e &=& (x-y-z+w)/8 + O(e^{-\bbeta}) \\
f &=& (-x+y+z-w)/8 + O(e^{-\bbeta}) \\
\eeann
where 
\beann
x &=& \ln(e^{-a+b} + \frac{1}{2} e^{a-b}) \\
y &=& \ln(\frac{3}{2} e^{-a+b} + \frac{1}{2} e^{a-b}) \\
z &=& \ln(e^{-a-b} + \frac{1}{2}e^{a+b}) \\
w &=& \ln(\frac{3}{2} e^{-a-b} + \frac{1}{2} e^{a+b})
\eeann
We will choose $a,b$ to make the coefficients of
$\s_0 s_{-1}$ and $\s_0 s_{-2}$ in \reff{fpeerr} almost vanish.
We let $a_0,b_0$ be the solution to the equations 
$c - \bbeta -a =0$ and $d-b=0$ if we ignore the $O(e^{-\bbeta})$ terms.
So
\bea
a_0 &=& (-2*\ln(1.5)+x-y+z-w)/8   \nonumber \\
b_0 &=& (-2*\ln(1.5)-x+y-z+w)/8 
\label{abfpe}
\eea
where $x,y,z,w$ are computed using $a_0,b_0$ for $a,b$. 
Note that these equations are independent of $\bbeta$.
With this choice,
\beann
||F(c_0+\hat{H_0}) - c_0|| = |e| e^{\mu} + |f| e^{\mu} + O(e^{-\bbeta})
\eeann

The solution to \reff{abfpe} is approximately given by
\beann
a_0 \approx -0.18019161, \quad b_0 \approx -0.02254094 
\eeann
Here and in the following we use $\approx$ to indicate that we are giving
numerical approximations that are accurate to eight decimal places.
For a fully rigorous proof we should use interval arithmetic for these
computations, but we have not done so. 
For this choice of $a,b$ we have
\beann
e\approx  0.00078810, \quad f\approx -0.00078810
\eeann
So 
\beann
||F(c_0+\hat{H_0}) - c_0|| \approx 0.00157619 e^{\mu}  + O(e^{-\bbeta})
\eeann

The expectations $<\s(A)>_{c_0}$ are functions of $\s_1,s_0,s_{-1},s_{-2}$. 
For large $\bbeta$ they are independent of $\bbeta$ up to terms of order
$e^{-\bbeta}$. Their computation is straightforward but tedious. One can give
explicit but complicated expressions in terms of $a$ and $b$. We only give
the numerical results for $a=a_0$ and $b=b_0$.
\beann
<\s_0>_{c_0} &\approx& 
  0.77632018  s_0 
+ 0.22367982  s_{-1} 
- 0.03496197  \s_2 
+ 0.03496197  s_{-1}  s_0  \s_2
\\
&-& 0.00777073  s_{-2} 
+ 0.00777073  s_{-2}  s_{-1}  s_0 
- 0.00087107  s_{-2}  s_0  \s_2 
+ 0.00087107  s_{-2}  s_{-1}  \s_2 
\\
<\s_1>_{c_0} &\approx& 
  0.69811964  s_0 
+ 0.30188036  \s_2 
- 0.03145297  s_{-1} 
+ 0.03145297  s_{-1}  s_0  \s_2
\\
&+& 0.00114994  s_{-2} 
- 0.00114994  s_{-2}  s_0  \s_2 
- 0.00114994  s_{-2}  s_{-1}  s_0 
+ 0.00114994  s_{-2}  s_{-1}  \s_2 
\\
<\s_0 \s_1>_{c_0} &\approx& 
0.47443982
+  0.26691838  s_0  \s_2
+  0.19222684  s_{-1}  s_0 
+  0.06641495  s_{-1}  \s_2
\\
&-& 0.00662078  s_{-2}  s_0 
+ 0.00662078  s_{-2}  s_{-1} 
- 0.00202101  s_{-2}  \s_2 
+ 0.00202101  s_{-2}  s_{-1}  s_0  \s_2 
\eeann
Thus by \reff{ddef}
\beann
\D(c_0)\approx \min\{
&& 0.07166608 + 1.01554146 e^{-\mu},  \, 
0.33563322 + 0.73187250  e^{-\mu}, \\
&& 0.33737535  e^{-\mu} + 0.67990824  e^{-2 \mu}
\}
\eeann

There are many choices of $\mu$ for which the hypothesis of the theorem is
satisfied for small $\epsilon$.
For example, with $\mu=1.0$ we have $||F(c_0)-c_0||\approx 0.00428454$ 
and $\D(c_0)\approx 0.60487407$.
The infimum in \reff{thmhyp} is approximately $0.25088335$
when $\epsilon=0$ and so it is less than $1$ for small $\epsilon$. 

\section{ Proof of LRO by reflection positivity}
\label{lro_sect}

In this section we sketch a proof that the Hamiltonian \reff{myham}
has long range order if $\epsilon>0$ and $\bbeta$ is large enough
(depending on $\epsilon$).

\begin{proposition} 
Let $1 < \alpha < 2$.
Define for positive $\bbeta$ and $\epsilon$ 
  \beann
  H = - (\bbeta-\epsilon) \sum_i \s_i \s_{i+1} 
  - \epsilon \sum_{j<k} \frac{\s_j \s_k}{|j-k|^\alpha} 
  \eeann
  For all $\epsilon>0$ there exits $\bbeta_0$ which depends on $\epsilon$ and
  $\alpha$ such that there is long range order if $\bbeta > \bbeta_0$. 
\end{proposition} 
For consistency with \cite{frohlich1978phase} we have included the 
nearest neighbor terms in the second sum. Note that the Hamiltonian above is
the same as \reff{myham}.
We emphasize that nothing in this section is new.
This result follows from the methods
in \cite{frohlich1978phase}. However, finding exactly the ingredients
needed for our case in \cite{frohlich1978phase} takes a bit of time,
so we highlight these ingredients here. We assume familiarity with
the general techniques of \cite{frohlich1978phase}.

For a positive integer $m$ we let
\bea
\Lambda=\{1-m,2-m, \cdots -2, -1, 0, 1, 2, \cdots, m\}
\label{lambda_def}
\eea
We impose periodic boundary conditions so $m+1$ means $1-m$ and $-m$ means $m$. 
Up to a constant, the Hamiltonian can be rewritten as 
\beann
H=  \frac{1}{2} \sum_{j, k \in \Lambda}
[(\bbeta - \epsilon) N_{j,k}  + \epsilon J_{j,k}] (\s_j - \s_k)^2
\eeann
where $J_{j,k}= |j-k|^{-\alpha}$ and $N_{j,k}$ is $1$ if $j,k$ are nearest
neighbors and $0$ if they are not.
To make this periodic we take $J_{j,k}$ to be 
\bea
J_{j,k} = \sum_{n=-\infty}^\infty \frac{1}{|j-k+2mn|^\alpha}
\label{periodicj}
\eea
The Hamiltonian depends on $m$, but
we do not make this dependence explicit in the notation.

Define 
\beann
g_m(p) = <\hat{\s}_p \hat{\s}_{-p} >
\eeann
where
\beann
\hat{\s}_p =\frac{1}{\sqrt{2m}} \sum_{j \in \Lambda} e^{ipj} \s_j
\eeann
and $p$ is an element of the dual lattice
$\Lambda^* = \{0,\frac{\pi}{m},\frac{2\pi}{m},\cdots,\frac{(2m-1)\pi}{m}\}$.
Here $< \quad>$ is the Gibbs measure on $\Lambda$. 
Let
\beann
E(p) = \frac{1}{2} \sum_{j \in \Lambda} (1-e^{ipj}) 
[(\bbeta - \epsilon) N_{j,0}  + \epsilon J_{j,0}] 
= (\bbeta - \epsilon) (1-\cos p) + \frac{\epsilon}{2} R(p) 
\eeann
where 
\beann
R(p) = \sum_{n \in \Lambda: n \neq 0 } J_{0,n} (1-\cos(pn))
\eeann
The infrared bound says that for non-zero $p$ 
\bea
g_m(p) \le \frac{1}{2 E(p)}
\label{ir}
\eea
This bound will follow from the Gaussian domination bound which
we now state. For a real valued function $h_i$ on $\Lambda$ define
\bea
Z(h) = < \exp(-\frac{1}{2} \sum_{j,k}
[(\bbeta - \epsilon) N_{j,k}  + \epsilon J_{j,k}] |\s_j-\s_k-h_j+h_k|^2) >_0
\label{zdef}
\eea
where $< \cdots >_0$ is the sum over the spins $\s_i$ with $i \in \Lambda$,
normalized so that $<1>_0=1$. 
In the above the sums over $j,k$ are over $\Lambda$. 
Gaussian domination says that for all $h$ 
\bea
Z(h) \le Z(0)
\label{gd}
\eea

\begin{proposition}
The Gaussian domination bound \reff{gd} implies the infrared bound \reff{ir}.
\end{proposition}

\medskip
\no {\bf Proof:} 
Take $h_j = \delta e^{ipj}$ and expand the Gaussian domination bound to second
order in $\delta$. See \cite{frohlich1978phase} for details. \qed

\medskip

Since we are in one dimension, the integral of $1/(1-\cos p)$ near $0$
diverges. The $R(p)$ term will act as a sort of regularizer.
\begin{proposition} 
  \bea
  \sup_m \, \frac{1}{2m} \sum_{p \in \Lambda^*} \frac{1}{R(p)}< \infty
  \eea
\end{proposition} 

\no {\bf Proof:} 
This can be found in \cite{frohlich1978phase}.  We give a proof since it is
central to the proof of LRO. 
As $m \rightarrow \infty$ the normalized sum on $p$ converges to a normalized
integral over $p \in [-\pi,\pi]$. Since the integrand is even, we can take the
intergral just over $[0,\pi]$. So we need to show 
\bea
  \int_{0}^\pi \frac{dp}{R(p)}< \infty
\eea
We get a lower bound on $R(p)$ by restricting the sum
on $n$ to those with $\frac{\pi}{2p} \le n \le \frac{\pi}{p}$. 
For such $n$ we have $pn \in [\pi/2,\pi]$ so
$1-\cos(pn) \ge 1$.
We have $J_{0,n} = n^{-\alpha} \ge p^\alpha/\pi^\alpha$. The number of $n$ in the
restricted range goes as $\pi/(2p)$ as $p \ra 0$. So 
\beann
R(p) \ge c p^{\alpha-1}
\eeann
for some constant $c$. Since $1<\alpha<2$
the integral of $1/p^{\alpha-1}$ is finite. \qed

If we fix $\epsilon$, then by the dominated convergence theorem,
\beann
\lim_{\bbeta \ra \infty} \int \frac{dp}{E(p)} = 
\lim_{\bbeta \ra \infty} \int \frac{dp}
    {(\bbeta-\epsilon) (1-\cos(p)) + \frac{\epsilon}{2} R(p)} 
= 0
\eeann
So by the usual argument, for large $\bbeta$ there must be an atom at the
origin in $g_{\Lambda}(p)$ which shows there is long range order.

The key to proving the Gaussian domination bound is  the following proposition.

\begin{proposition}
  We write $Z(h)$ as 
  $Z(h_{1-m},h_{2-m},\cdots, h_{-1},h_0;h_1,h_2,\cdots,h_m)$
  where the semicolon helps us see the reflection point at $1/2$. Then
  \beann
&&  |Z(h_{1-m},h_{2-m},\cdots, h_{-1},h_0;h_1,h_2,\cdots,h_m)|^2 \le \\ 
  && |Z(h_m,h_m-1,\cdots, h_2,h_1;h_1,h_2,\cdots,h_m)| 
  |Z(h_{1-m},h_{2-m},\cdots, h_{-1},h_0;h_0,h{-1},\cdots,h_{1-m})|
\eeann  
\label{rpbound}
\end{proposition}

One can use this proposition to show that the maximum of $Z(h)$ is attained
by an $h$ in which the $h_i$ are all equal. This of course is the same
as $Z(0)$, and so proves Gaussian domination.
The proof of the above can be found in \cite{frohlich1978phase}.
The essential ingredient is that $N_{j,k}$ and $J_{j,k}$ are reflection positive.
We sketch the proof that $J_{j,k}$ is reflection positive
and refer the reader to 
\cite{frohlich1978phase} for the rest of the proof of the proposition.

The reflection is about $1/2$. So
terms that cross this reflection point are $J_{j,1-k}$ where
both $j$ and $k$ range from $1$ to $m$. 
Reflection positivity follows from an integral representation for these $J$:
\bea
  J_{j,1-k} = \int_0^1 [\lambda^{j+k-1} (1-\lambda^{2m})^{-1} +
   \lambda^{m-j+m-k+1} (1-\lambda^{2m})^{-1}] \, \mu(d \lambda)
\label{intrep}
\eea
where $\mu$ is a positive measure on $(0,1)$. 
To derive this representation we start with 
\bea
\int_0^\infty e^{-nx} \, x^{\alpha-1} \, dx = \frac{\Gamma(\alpha)}{n^\alpha}
\eea
Using a change of variables $\lambda=e^{-x}$, this shows there is a positive
measure $\mu(d \lambda)$ on $(0,1]$ such that
\bea
n^{-\alpha} = \int_0^1 \lambda^n \, \mu(d \lambda)
\eea
By taking into account the constraints $1 \le j,k \le m$ we can get rid of
the absolute values in the definition of $J_{j,1-k}$ in \reff{periodicj}
and then sum the two resulting geometric series to obtain \reff{intrep}.

\bigskip

\no {\bf Acknowledgments:} This research was supported in part by NSF grant
  DMS-1500850.

\bibliography{rg_1d_power_law}{}
\bibliographystyle{plain}

\end{document}